Research Paper

# The Impact Of Country Of Origin In Mobile Phone Choice Of Generation Y And Z

Szabolcs Nagy

Corresponding author: marvel@uni-miskolc.hu

Faculty of Economics University of Miskolc, Hungary

## About the author

Szabolcs Nagy is an associate professor and the head of Department of Marketing Strategy and Communi-cation at Faculty of Economics, University of Miskolc, Hungary. He received his Ph.D in Management and Organizational Sciences in 2005. He has special interest in both technology and digital marketing. With twenty years of experience in marketing, he has been very active in business as a consultant with more than thirty successful projects.

## abstract

Mobile phones play a very important role in our life. Mobile phone sales have been soaring over the last decade due to the growing acceptance of technological innovations, especially by Generations Y and Z. Understanding the change in customers' requirement is the key to success in the smartphone business. New, strong mobile phone models will emerge if the voice of the customer can be heard. Although it has been widely known that country of origin has serious impact on the attitudes and purchase decisions of mobile phone consumers, there lacks substantial studies that investigate the mobile phone preference of young adults aged 18–25, members of late Generation Y and early Generation Z. In order to investigate the role of country of origin in mobile phone choice of Generations Y and Z, an online survey with 228 respondents was conducted in Hungary in 2016. Besides the descriptive statistical methods, crosstabs, ANOVA and Pearson correlation are used to analyze the collected data and find out significant relationships. Factor analysis (Principal Component Analysis) is used for data reduction to create new factor components. The findings of this exploratory study support the idea that country of origin plays a significant role in many respects related to young adults' mobile phone choice. Mobile phone owners with different countries of origin attribute cruicial importance to the various product features including technical parameters, price, design, brand name, operating system, and memory size. Country of origin has a moderating effect on the price sensitivity of consumers with varified net income levels. It is also found that frequent buyers of mobile phones, especially US brand products, spend the most significant amount of money for their consumption in this aspect.
Keywords:mobile phone, product choice, country of origin, Generation Y, Generation Z

## Introduction

Mobile phone use has been soaring over the last decade. People with different ages show preference in their mobile phone choice. Some of them are even addicted to their mobile phone use, which occupies a large portion of their daily consumption. In Hungary, the number of mobile phone subscriptions per thousand inhabitants was 1,189 in 2015 (KSH, 2017a). This implies that one individual subscribes more than one mobile phone throughout the country. However, research that explores the mobile phone choice of people of different ages has been lacking.

The change of customer requirements stems from generation gaps. This understanding has utmost importance in business. Without sufficient recognition of the changes in customers' preference, Nokia's market share decreased from 50.9% in the 4th quarter of 2007 to around 3.0% in 2013 just in seven years (Statista, 2017). Nokia's case is the best real-life example of how a strong, well-established market-leader brand can be ruined due to the lack of understanding the needs of the new generations. Generation Y, also called Echo Boomers or the Millenniums, born between 1977 – 1994, are technology-wise people. They possess strong digital knowledge, tend to be less brand loyal, and are resistant against marketing and advertising (Bajner, 2012). This is also applicable when it comes to their mobile phone choice.

Generation Z, born between 1995 – 2012, are the children of the Digital Age. They were born into a highly diverse environment, and rapidly spreading e-solutions play a fundamental role in their daily life. Smartphones have a leading position in their information acquisition and social network, and function as their indispensable partner. (Ozkan and Solmaz, 2015). They are the most Internet savvy generation ever and have more digital literacy than Genernation Y (Socialmarketing, 2017).

According to T´obi and T¨or˝ocsik (2013), there are several significant differences between Generation Y and Generation Z. Generation Z need constant and instant access to the internet, and make sure that they are always connected to the outside world online. For them, smartphones are not only a status symbol of their identity, but also a device to maintain their everyday activities. They are convinced that people without sufficient online connection are less competitive in nowadays turbulent society.

Features of mobile phones have serious impact on consumers' acceptance and their popularity. Kekolahti et. al (2016) use a systematic longitudinal analysis to anayalyze the popularity of mobile phones. They discovered that the operating system was undoubtedly the only feature with an increasing strength in predicting the popularity of mobile phones between 2004 and 2013, while other features, i.e. display, camera, etc., were more or less important or have structural breaks in their influence over time. Head and Ziolkowski (2012) identify two distinct segments of mobile phone users among university students: instant communicators and communicators/information seekers. The communicators/information seekers perceive their mobile phones as a utilitarian tool, while the instant communicators hold a more hedonic perspective for their mobile phone. They have a more positive attitude towards mobile phone and enjoy using this convenient device.

The variable country of origin refers to the country of the brand owner of a mobile phone, which has a direct impact on a consumer's perception of the brand. This influence can be both negative and positive (Cateora and Graham, 1999). The perceived country-of-origin functions as a decisive factor in the consumers' attitudes and intentions in purchasing any products and brands (Hanzaee and Khosrozadeh, 2011).

For baby boomers – people born between 1946 – 1965, as well as the majority of Generation X – people born between 1966 – 1976, a mobile phone is a perfect choice for text messaging. However, Generation Y – people born between 1977 – 1994, demand more advanced mobile phone features, such as larger capability for browsing the internet, taking photos, playing games, etc. Therefore, comprehensive analysis of mobile phone choice of people aged 18–25 is timely and relevant to the promotion of mobile phone consumption. This age-group includes two decisive generation cohorts, the members of late Generation Y and early Generation Z.

This research aims to analyze the mobile phone choice of people aged 18–25 in Hungary with special regard to country of origin as a moderator variable. This study focuses on the following three questions: (1) how much is each mobile phone feature taken into account during the purchase process; (2) what brand names and country of origins are preferred by young adults; (3) what information sources are important and what webshops are favoured. Moreover, this study also investigates the typical use of mobile phones and people's willingness to pay to create a demand curve. Country of origin is assumed to be an important factor on young people's product choice.

## Methodology

For data collection, an online questionnaire survey was conducted in September, 2016 in Hungary. A valid sample of 228 respondents is used for this analysis. Considering the size of the population in Hungary with 974,655 people aged 18-25 in 2016 (KSH 2017b), the sample size of 228 respondents resulted in a 6.49% confidence interval at the 95% confidence level.

The sample was representative with respect to the gender variable. 51.3% of respondents are males, 48.7% are females. 48.4% of the respondents are from Generation Z, born between 1995 and 1998. 51.6% of the respondents were born between 1991 and 1994, so they represent Generation Y. As far as other demographical variables are concerned, the sample was not representative. As for education, only very few of the respondents finished the basic-level education (1.8%). Most people have higher education experience with a university or college degree (54.4%) or finished secondary education with a certificate (43.9%). Most of the respondents live in a major town/city of the country (39.0%), or in a medium-level town (27.2%). 22.4% of the respondents live in a village, and 11.4% of them reside in the capital (Budapest).

Descriptive statistical methods, in a more precise term, frequencies are used to analyze means and median. Crosstabs, ANOVA and Pearson correlation are used to find out significant relationships among variables. Factor analysis (Principal Component Analysis) is used for data reduction to create factor components of variables.

## Findings and Discussion

The findings of the research in this section are presented based on the 4P concept as below: Product, Place Promotion and Price.

### Product

Most people aged 18 25 (82.9%) prefer strong, well-established brands, therefore the ratio of weak brands occupies only 17.1%. As far as country of origin is concerned, brands from South-Korea (32.5%) are the most popular among people within this age group in Hungary, followed by Chinese mobile phones (28.9%) and US brands (18.9%). The market shares of EU brands (11%) and Japanese brands (8.8%) are rather low.

Country of origin is defined according to the country where the brand owner/company has registered their product. Users own mobile phones whose brands were registered in different countries, therefore, the "current country where the brand is owned variable has been transformed into a new variable, named "country of origin . For example, Apple, Microsoft and Caterpillar are considered as US brands; Nokia is the one and the only EU brand and Sony is the only Japanese brand. Samsung and LG are considered as South Korean brands, and at last but not at least, all the other brands including Huawei, Lenovo, Alcatel, One Plus, etc. are categorized as Chinese mobile phone brands.

Brands are divided into strong and weak categories based on the perceived power of those brands in Hungary. Apple, Caterpillar, Huawei, LG, Microsoft, Nokia, Samsung and Sony are considered as strong brands, while other brands are weak brands because they hold relatively weak brand awareness to consumers. In this latter group, all of the products are Chinese brands. US, EU, Japanese and Korean brands are 100% strong brands. 40.9% of young adults who own a Chinese mobile phone buy products with a strong brand name, and 59.1% of them buy a weak brand. As far as current brands owned are concerned, the TOP3 brands are Samsung (27.6%), Apple (17.1%) and Huawei (11.8%). Nokia (11.0%) and Sony (8.8%) are also found to be popular brands. LG ranked the sixth among all (4.8%), followed by Lenovo (3.5%), Alcatel (3.1%) and One Plus (1.8%). Other current brands owned by Hungrian users represent 10.5% in total, including Doogee (1.3%), Microsoft (1.3%), Prestigio (1.3%), Xiaomi (1.3%), HTC (0.9%), Navon (0.9%), Vodafone (0.9%), Blackview (0.4%), Caterpillar (0.4%), Cirrus (0.4%), Gsmart (0.4%), Meizu (0.4%) and ZTE (0.4%).

(1) Factors Related to Consumers' Mobile Phone Choice

    A five-point Likert scale is used to meausre the means for the various parameters related to the choice of consumers when making decisions in their new mobile phoen.   Score 5 represents extremely important and score 1 represents least important.  As a result, when it comes to purchasing a new mobile phone, price (4.27) turns out ot be the most important feature, followed by battery capacity (4.19), technical parameters (4.15) and memory size (4.13).  Parameters including operating system (3.96), warranty period (3.94), design (3.89) and display size (3.86) show average importance, whereas color (3.32) and brand name (3.23) are the least important features.

    In order to find out relationships of variables determining the mobile phone choice, the rotation method of varimax with Kaiser Normalization is used for principal component analysis. Kaiser-Meyer-Olkin Measure of Sampling Adequacy (KMO) was 0.769. Rotated component matrix (Table 1) shows that all variables are grouped into three factor components, with an explained cumulative variance at the level of 0.614. The first component (26.1% of variance), called 'Technical components', includes operating system (OS), memory size, technical parameters and display size. The second component (20.4% of variance), called 'Appearance component', includes visible features which can be perceived by others, such as design, color and brand. The third component ( 14.8% of variance) contains mixed features, such as battery capacity, price and warranty period.

    The results of Pearson correlation also support the above findings.  As shown in Table 2, there is a strong correlation between design and color (0.70).  A moderate correlation is found between operating system and technical parameters (0.6), technical parameters and memory size (0.57), memory size and operating system (0.53).  Weak correlations are found between memory size and battery capacity (0.49), battery capacity and warranty (0.43),  technical parameter and display size (0.37),  technical parameters and design (0.35), warranty and memory size (0.35), operating system and brand name (0.34), technical parameters and brand name (0.33) as well as operating system and design (0.33).  There are also weak correlations between brand name and color (0.31), battery capacity and display size (0.31) and technical parameters and warranty (0.30).

    In order to find out the significant differences among the important features of mobile phones with

Table 1   Rotated Component Matrix of Product Features

|  | Technical Component (26.1%) | Appearance Component (20.4%) | Other Components (14.8%) |
|---|---|---|---|
| Operating System | .838 | .177 | -.072 |
| Memory Size | .780 | .082 | .303 |
| Technical Parameters | .754 | .228 | .119 |
| Display Size | .526 | .363 | .268 |
| Design | .197 | .887 | .063 |
| Color | .056 | .884 | .036 |
| Brand Name | .354 | .504 | -.120 |
| Battery Capacity | .369 | .021 | .673 |
| Price | -.195 | .008 | .649 |
| Warranty Period | .341 | .000 | .634 |

different countries of origin, ANOVA is used for further analysis. It is found that there are six features which indicate the significant differences among mobile phones with different countries of origin: technical parameters, price, design, brand name, operating system, and memory size (see Fig. 1). In the case of US models (Apple, Microsoft and Caterpillar), the operating system is the decisive factor, but brand names also

Table 2   Pearson Correlations of Mobile Phone Features

|  | TP | P | C | D | BN | OS | DS | MS | BC | W |
|---|---|---|---|---|---|---|---|---|---|---|
| TP | 1 | .083 | .258** | .353** | .333** | .604** | .369** | .567** | .260** | .302** |
| P | .083 | 1 | -.105 | -0.023 | 0.003 | -0.039 | .086 | .025 | 0.095 | 0.112 |
| C | .258** | -.105 | 1 | .698** | .305** | .196** | .280** | .182** | 0.125 | 0.079 |
| D | .353** | -.023 | .698** | 1 | .394** | .329** | .425** | .257** | .136* | 0.11 |
| BN | .333** | .003 | .305** | .394** | 1 | .341** | .212** | .231** | 0.058 | 0.112 |
| OS | .604** | -.039 | .196** | .329** | .341** | 1 | .480** | .529** | .207** | .216** |
| DS | .369** | .086 | .280** | .425** | .212** | .480** | 1 | .473** | .308** | .229** |
| MS | .567** | .025 | .182** | .257** | .231** | .529** | .473** | 1 | .493** | .348** |
| BC | .260** | .095 | .125 | .136* | 0.058 | .207** | .308** | .493** | 1 | .434** |
| W | .302** | .112 | .079 | 0.11 | 0.112 | .216** | .229** | .348** | .434** | 1 |

TP:Technical Parameters, P:Price, C:Color, D:Design, BN:Brand Name, OS:Operating System
DS: Display Size, MS:Memory Size, BC: Battery Capacity, W;Warranty

Notes:    * Correlation is significant at the 0.05 level (2-tailed).

          ** Correlation is significant at the 0.01 level (2-tailed).

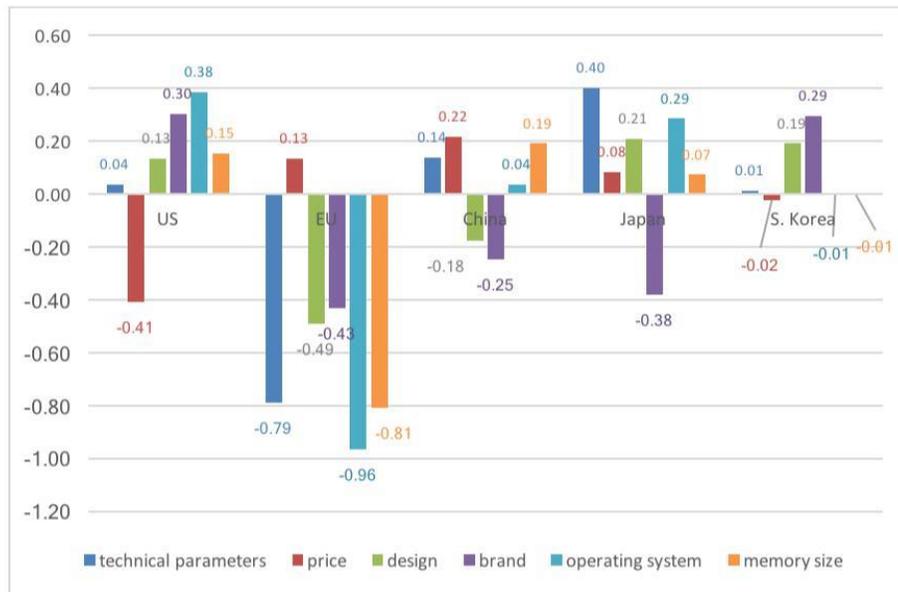

Fig. 1   Differences Among Means of Important Features According to Country of Origin

have significant impact on consumers' decisions, along with memory size and design. For US models, price plays a significantly weak role in the purchasing decision-making process than in the case of models from other countries. For buyers of South Korean mobile phones (LG and Samsung), brand name and design are the significantly important factors which indluence consumers' final choice. Comsumers who prefer Japanese mobile phones (Sony) attribute higher importance to technical parameters, design and operating system (OS), but surprisingly they pay much less attention to brand names. That is to say, Hungarian consumers believe that all Japanese mobile phone products maintain the same technical level. Price, memory size and technical parameters are important factors attracting Chinese mobile phones consumers. For people aged 18–25 who use EU mobile phones (Nokia), price is the most attractive feature and has made the strongest impact on their choice, while all the other features turn out to be significantly less important.

The survey also shows that Hungarian young adults usually replace their mobile phones in every two (36.8%) or three years (24.1%). However, some (18.0%) of the young people aged 18–25 buy a new mobile only when the old one is not working anymore. 16.6% of the young people purchase a new mobile phone in every four years, or even use their old models for a longer period of time (2.1%). Only 5.9% of the young people can be considered as frequent buyers of mobile phones. They change their mobile phones into new models within twelve months.

(2) Functions Popular with Mobile Phone Users

A five-point Likert scale is also employed to measure the frequency young adults make use of the various functions of their mobile phones, with score 1 representing never and score 5 representing almost always. The data shows that young people most frequently use their mobile phones to make phone calls (4.55) and browse the internet (4.43). Enjoying the various apps (3.93) and listening to music (3.72) are also very popular activities among young people. Taking photos & videos (3.34) and playing games (3.17) are the least frequently used with mobile phones. The different functions related to mobile phone use are classified into two groups of factors as a result of a Principle Component Analysis, with Kaiser-Meyer-Olkin Measure of Sampling Adequacy = 0.776, Bartlett's Test of Sphericity Approx. Chi-Square 328.958, df = 15, Sig. = .000. and V ariance = 0.613. Table 3 shows that the first component (44.5 % of variance) includes all functions except one variable – making phone calls. The "making phone calls variable makes up the second component (16.8% of variance), which supports the idea of the existence of communicators/information seekers identified by Head and Ziolkowski (2012) as mentioned above.

Table 3   Rotated Component Matrix of Mobile Phone Use

|  | Components | |
|---|---|---|
|  | All Functions Except Phone Calls (44.5%) | Phone Calls (16.8%) |
| Browsing Internet | .844 | -.005 |
| Enjoying Apps | .832 | .047 |
| Taking Photos Videos | .742 | -.083 |
| Listening to Music | .629 | .026 |
| Playing Games | .563 | .129 |
| Making Phone Calls | .031 | .990 |

## Place

The term place in this study refers to where young people purchase their mobile phones. Only 17.6% of young adults don't make use of online mobile phone shops, while the rest (82.4%) all indicate that they have their preferred webshops. The most preferred webshop among them is emag.hu (14.5%), which is an online plaza. It is followed by arukereso.hu (13.2%), a price comparison web page and Telekom.hu (11.8%), the biggest mobile service provider in Hungary. 7.9% of the young people preferred jofogas.hu, which is one of the biggest online classified ad sites in Hungary. Argep.hu, another price comparison site, is preferred by 6.6% of young adults. While in the case of other webshops including edigital.hu, an online plaza, a significant slump in the number can be seen. Surprisingly, only 2.2% of the young adults stated that they prefer the webshop of Media Markt, which is the largest retailer for electronics in Europe and an established brick and click company. Apple.hu was preferred by only 1.8% of the young adults.

As far as international e-tailers are concerned, eBay (11.0%), Alibaba.com (1.8%) or Tinydeal.com (0.4%) are also possible options for Hungarian consumers of mobile phones, but most of them (96.7%) have a relatively strong preference for the local shops. A crosstab analysis reveals that owners of US mobile phone products has a significantly weak preference for shopping online (adjusted residual=2.0).

Table 4  Rotated Component Matrix of Mobile Phone Use

|  | Professional Component (28.82%) | Friend & Family Component (23.14%) | Online Component (15.95%) |
|---|---|---|---|
| Store | .795 | .107 | .011 |
| Service Providers | .790 | .115 | .121 |
| TV | .710 | .158 | -.044 |
| Friends | .110 | .893 | .099 |
| Family Members/Relatives | .184 | .878 | -.050 |
| Internet | -.071 | .023 | .925 |
| Magazines | .453 | .033 | .482 |

## Promotion

Promotion stands for the efforts of mobile phone companies in promoting sales. It includes advertisement and all other activilites they hold to make their products better known. A five-point Likert scale is used to show the importance of various sources users appeal to for information regarding mobile phones. Score 1 stands for not important, whereas score 5 represents very important.

The survey suggestsnternet is the Internet is the only very important source of information for consumers when they purchase new mobile phones (4.54). Friends (3.14) have an average influencing power, while family members/relatives (2.71), store personnel (2.51) and mobile service providers (2.38) contribute much less to their fianl decision. Magazines (1.91) and TV (1.68) show no perceptible impacts in this regard.

There is no significant difference regarding the information sources between the items within the category of country of origin. Therefore, it can be concluded that consumers of mobile phones, regardless of country of origins of their phones, are more likely to search for information on the internet. Promotions on the internet is the most effective in improving the reputation of a mobile phone product.

In order to find out more about the structure of information source variables, principal component analysis is used. As a result, K M O = 0.656, Bartlett's Test of Sphericity Approx. C hi − Square = 300.371, df = 21, Sig. = 0.000. As shown in Table 4, all variables for information source are divided into three factor components. The first group of components, named professional components, comprises store personnel, service providers, TV and magazines (28.82% of variance). Friends and family members/relatives make up the second component (23.14% of variance), while internet and magazines compose the online factor component (15.95% of variance).

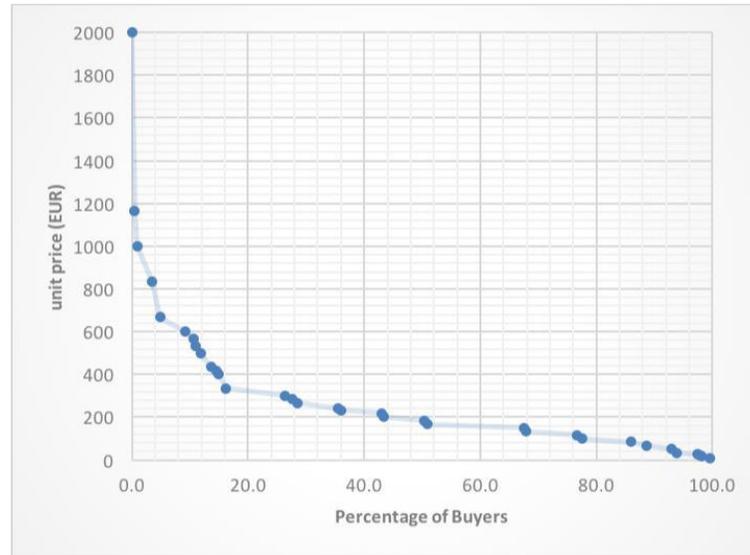

Fig. 2    Demand Curve for Mobile Phones

## Price

In order to obtain information about the willingness of young people to pay for a new mobile phone, a simple question was asked – 'How much is the maximum amount of money you are willing to pay for a new mobile phone?' The mean of the answers to this question is 79,144 HUF (~264 EUR), the median is 60,000 HUF (~200 EUR). A significant, but very weak relationship is found between the strength of the brand name and the maximum amount of money for a mobile phone (ANOVA Sig. = 0.04, η2 = 0.019). The result also shows that young adults prefer to buy a mobile phone with a strong brand (83,672 HUF, ~279 EUR) to a weak one (57,205 HUF, ~191 EUR) though they have to pay more money.

A demand curve is calculated. Based on the replies to the above question. The maximum amount of money consumers are willing to pay for a new mobile phone is 600,000 HUF (~2,000 EUR), but only 0.44% of people aged 18–25 show their willingness to do so. At the other extreme, 3,000 HUF (~10 EUR) is the lowest price limit for a new mobile phone product. If the price of a new mobile phone is below this amount, 100% of the young adults say they would buy it, provided that the product meets other expectations.

Fig. 2 shows that there are four major breaking points in the demand curve. To young consumers, the most important psychological limit is 50,000 HUF (~167 EUR). If the price of a mobile phone reaches this amount, it causes an immediate 16.7% drop in its demand. The price threshold in this case is 49,999 HUF (169.99 EUR). The second significant breaking point is 100,000 HUF (~333 EUR), which causes an immediate 10.1% drop in demand. The third one is 40,000 HUF (~133 EUR) causing a 8.8% loss of customers, while the fourth one is 30,000 HUF (~100 EUR) with a 8.3% demand cut. It can also be seen that 60,000 (200 EUR), 70, 000 HUF (~233 EUR) and 80,000 HUF (~267 EUR) are also important price levels, each causing a 7.0% drop in the total demand.

Based on the above findings, customer can be divided into four segments regarding to the price they are willing to pay for a mobile phone. For the segment of customers who are the most sensitive to price, a very basic phone version should be offered at a very cheap price. The suggested price is below 29,999 HUF (~99.99 EUR). More features can be added to this basic model when the segments of customers who show more buying potential and want to enjoy the various functions with their phones.

It is surprising that there is a weak, but significant relationship (ANOVA, sig = 0.00 and $\eta2$ = 0.146) between the maximum amount of money spent on a mobile phone and the replacement period. Those who purchase a new mobile every year, spend the most for a new device (165,000 HUF, ~550 EUR). Young adults who replace their phones once every two years are willing to pay much less (98,000 HUF, ~327 EUR). People replacing their devices once every three years pay an average of up to 73,000 HUF (~243 EUR), while those who buy a new phone every four years spend only 55,000 HUF (~183 EUR). Even less, 45,000 HUF (~150 EUR) is paid when the mobile phone is replaced every five years. In the case of an unintentional replacement, e.g. when the old mobile phone goes wrong and doesn't work anymore, an average amount of 50,000 HUF (~167 EUR) is spent for every new model.

There is a weak, but significant relationship between the the brands and the maximum amount of money spent for a new mobile phone as the results of an ANOVA analysis (sig = 0.16 and $\eta2$ = 0.158). Young adults pay the most for an Apple (129,230 HUF, ~431 EUR), then for a One Plus (111,250 HUF, ~371 EUR) or a Samsung (87,206 HUF, ~291 EUR). HTC, Huawei, Navon, Sony and ZTE can be found in the price range of 70–80,00 HUF (~233–258 EUR). Alcatel, LG, Xiaomi, Blackview and Cirrus are brands usually bought with a price of 50-60,000 HUF (~167–212 EUR). Lenovo, Nokia, Gsmart and Meizu are brands with 40-50,000 HUF (~133-163 EUR). Prestigio and Microsoft are in the price range of 30–40,000 HUF (~111–122 EUR). Under this price range, Vodafone and Doogee can be found. People pay the least for a Caterpillar (20,000 HUF / ~67 EUR).

There is also a weak but significant relationship between country of origin and the maximum amount of money for a mobile phone (ANOVA, Sig. = 0.00, $\eta2$ = 0.101). Young adults pay the most (120,000 HUF, ~400 EUR) for US brands, a lot less (83,027 HUF, ~277 EUR) for Korean phones and even less for Japanese brands (70,750 HUF, ~236 EUR). It was found that young people are willing to pay only 65,015 HUF (~217 EUR) for Chinese mobile phones, and the least amount of money (41, 400 HUF, ~138 EUR) for EU brands.

Not surprisingly, it was found that there is a significant relationship between the net income per capita and country of origin of the brands (ANOVA, Sig. = 0.002, $\eta2$ = 0.073). The most affluent group of young adults with an average of 212,558 HUF (~708 EUR) net income per capita prefer a US brand (Apple, Microsoft and Caterpillar). The average net income per capita slumps in the next group of income category to 139,014 HUF (~464 EUR), where mobile phones from South Korea is more popular. Buyers of Japanese products (117,000 HUF, ~390 EUR) and EU brands (116,000 HUF, ~387 EUR) have a similar net income. Young adults with the lowest net income per capita (104,242 HUF, ~348 EUR) prefer Chinese brands.

Pearson correlation justifies the conclusion that there is a strong negative relationship (r = −0.390) between the maximum amount of money consumers pay for a new mobile phone and the importance of price. This correlation is significant at the 0.01 level (2-tailed). Therefore, the more important the price is to the consumers, the less amount of money are willing to spend on a new mobile phone.

Similarly, negative correlation is also found between the net income per capita and the importance of price (Pearson correlation is −0.25, significant at the 0.01 level (2-tailed) though weaker. With the increase in the net income, price becomes less important.

The average net income per capita (in family unit) is found to be 138,364 HUF (~461 EUR), with the median of 100,000 HUF (~334 EUR). No significant difference among demographic variables can be seen by country of origin of the mobile phone.

## Conclusions

The research presented in this paper seeks to provide a wider view of the mobile phone choice of young adults aged 18–25 in Hungary. It is found that the majority of people in this age-group, the members of the late generation Y and the early generation Z, usually purchase mobile phones of strong and established brands. However, surprisingly, when it comes to buying a new mobile phone, they consider price as the most important feature and brand name as the least. The more important the price is, the less amount of money they are willing to spend on a new mobile phone. Having considered preference patterns of consumers in Hungary, successful marketers should offer mobile phones with a strong battery, excellent technical parameters and sufficient memory at a reasonable price. Product features can be grouped into technical, appearance and economic factors, so each of them must be addressed when creating messages for the young adults.

It is an important managerial implication that when it comes to innovation, marketers must be aware of the fact that colour and design are strongly correlated, with operating system, technical parameters and memory size also having strong relationships. Significant differences are found between mobile phones with different countries of origin with regard to technical parameters, price, design, brand name, operating system, and memory size. US brand owners are the least price-sensitive customers. For them, the operating system and design are the decisive factors. On the other extreme, Chinese mobile phone owners are very price-sensitive due to their significantly lower net income. This also applies to Nokia's customers (EU brand). Young adults buy South-Korean mobiles for their preference of the brand and design, while Japanese brand (Sony) owners attribute significantly higher importance to technical parameters, design and OS, but surprisingly much less importance to the brand names.

It is also very suggestive to managers that Hungarian young adults usually replace their mobile phones every two or three years. Those who purchase a new mobile every year, are mostly US brand owners and tend to spend the most amount of money on a new device among all mobile phones users. Mobile phones are primarily used to make phone calls and browse the internet. Making phone calls makes up a separate factor component while all other use is involved in another. This indicates that making phone calls is the most significant among all its functions to young adults.

Most young adults have a preferred webshop. They like shopping online. However, owners of US mobile phones have significantly weaker preference for online shopping. They prefer visiting brick and mortar shops of mobile service providers. It is also found that internet is the only important source of information for mobile phone choice at the information search stage. Professionals, friends and family as well as online sources are the main categories which provide potential mobile phone buyers useful information.

This research supports the theory that country of origin has a significant impact on the maximum amount of money for a new mobile phone. It is found that young adults pay the highest amount for US brands (∼400 EUR), much less for Korean phones and even less for Japanese brands. It was also revealed that owners of Chinese mobile phones pay only about half of the money owners of US brands spend on their mobile phones. This ratio decreases to only 1/3 in the case of Nokia's customers (EU brand). Similar patterns can be seen in the net income of mobile phone owners. Most affluent young adults prefer US brands, while the poorest group prefer Chinese mobile phones. This study justifies that country of origin has a significant power of influence in many, but not every choice of the mobile phones.

It is also found in this article that four distinctive psychological limits exist regarding to the unit price of a mobile phone. Consequently, at least five major customer segments can be identified based on the price of each mobile phone in the market. Smart phone companies and retailers should take into account the impact of users' budgets when setting their shelf-prices and develop new models at different price levels.